\documentclass[twocolumn,prl,showpacs,nofootinbib]{revtex4}

\usepackage{graphicx}
\usepackage{dcolumn}
\usepackage{bm}
\def\hatn{\bm{\hat n}}

\def\VEV#1{\left\langle #1 \right\rangle}

\newcommand{\beq}{\begin{equation}}
\newcommand{\eeq}{\end{equation}}
\newcommand{\beqa}{\begin{eqnarray}}
\newcommand{\eeqa}{\end{eqnarray}}

\begin{document}

\title{Non-Uniform Cosmological Birefringence
and Active Galactic Nuclei}

\author{Marc Kamionkowski}
\affiliation{California Institute of Technology, Mail Code 350-17,
Pasadena, CA 91125}

\date{\today}

\begin{abstract}
Cosmological birefringence, a rotation by an angle $\alpha$
of the polarization of photons as they propagate over
cosmological distances, is constrained by the cosmic microwave
background (CMB) to be $|\alpha|\lesssim1^\circ$
($1\sigma$) out to redshifts $z\simeq1100$ for a rotation that
is uniform across the sky.  
However, the rotation angle $\alpha(\theta,\phi)$ may vary as
a function of position $(\theta,\phi)$ on the sky.  Here I
discuss how a position-dependent rotation can be sought in
current and future AGN data.  An upper limit $\VEV{\alpha^2}^{1/2} \lesssim
3.7^\circ$ to the scatter in the position-angle--polarization
offsets in a sample of only $N=9$ AGN already constrains the
rotation spherical-harmonic  coefficients to $(4\pi)^{-1/2}
\alpha_{lm}\lesssim 3.7^\circ$ and constrains the power spectrum
for $\alpha$ in models where it is a stochastic field.  Future
constraints can be improved with more sources and by analyzing
well-mapped sources with a tensor-harmonic decomposition of the
polarization analogous to that used in CMB polarization and weak
gravitational lensing.
\end{abstract}


\maketitle

{\it Introduction.} There is a very active quest to understand
dark energy \cite{darkenergy}, and quintessence models
\cite{quintessence} provide a promising set of effective theories.  
A pseudo-Nambu-Goldstone field provides an attractive
quintessence candidate, and such a field should have a coupling
to the Chern-Simons term of electromagnetism \cite{Carroll:1998zi}.
This coupling gives rise to cosmological birefringence (CB), a
frequency-independent rotation by some angle $\alpha$, of the
linear polarization of photons as they propagate
over cosmological distances
\cite{Carroll:1998zi,Carroll:1989vb}.  In the simplest models,
the rotation angle $\alpha$ is uniform across the sky, in which
case CB gives rise to parity-violating
TB and EB correlations in cosmic microwave background (CMB)
maps \cite{Lue:1998mq}.  Null searches for such correlations
now constrain the rotation over the redshift range
$0<z\lesssim1100$ to be $\alpha = -0.25^\circ\pm0.58^\circ$
\cite{Feng:2006dp}.

However, several recent papers have introduced quintessence
models in which the rotation angle $\alpha(\theta,\phi)$ varies
as a function of position $(\theta,\phi)$ on the sky
\cite{Pospelov:2008gg} and similar phenomena may arise in some
dark-matter models \cite{Gardner:2006za}.
Refs.~\cite{Kamionkowski:2008fp}
have described how to measure this rotation angle, as a function of
position on the sky, with the CMB, but the algorithm has not yet
been applied to data.  The WMAP satellite should be
sensitive (at $1\sigma$) to spherical-harmonic coefficients of
the rotation as small as $(4\pi)^{-1/2}|\alpha_{lm}|\sim 2.3^\circ$, for
$l\lesssim 400$, and the recently launched Planck satellite
should reach $(4\pi)^{-1/2}|\alpha_{lm}| \sim 0.07^\circ$ for
$l\lesssim 800$ \cite{Kamionkowski:2008fp}.

Radio
\cite{Carroll:1989vb,Nodland:1997cc,Loredo:1997gn,Wardle:1997gu,Leahy:1997wj,Carroll:1997tc}
and UV \cite{Cimatti:1993yc,Alighieri:2010eu} data on active
galactic nuclei (AGN) can
also be used to search for CB.
AGN are often elongated and polarized.
While AGN may be complicated objects, symmetry considerations
suggest that {\it on average} the mean polarization should be
aligned or perpendicular to the position angle of the source.
CB would, by rotating the polarization, give rise to a nonzero
mean offset between the position angles and polarizations
measured in a large number of sources
\cite{Carroll:1989vb,Nodland:1997cc,Loredo:1997gn}.
Likewise, if a detailed map of the intensity and polarization of 
an {\it individual} source can be made
\cite{Cimatti:1993yc,Wardle:1997gu,Leahy:1997wj}, then {\it on
average}, the intensity gradients and polarization within that
source should be aligned or perpendicular, if there is no CB.  
This more detailed analysis may allow competitive, or even
stronger, constraints to $\alpha$ from a smaller number of sources.

One precisely imaged radio source (3C 9) at a redshift
$z\simeq2$ constrains $\alpha=2^\circ \pm 3^\circ$ out to this
distance 
\cite{Leahy:1997wj,Kronberg}.  A stronger bound, $\alpha =
-0.6^\circ\pm1.5^\circ$, can be obtained from a larger number of
well-mapped radio sources, but only at smaller redshifts
\cite{Carroll:1998zi}.  A recent UV sample
\cite{Alighieri:2010eu} of eight
AGN at redshifts $z\gtrsim 2$ constrains $\alpha = -0.7^\circ
\pm 2.0^\circ$. In this Letter, I show that AGN can be used to
constrain the multipole moments $\alpha_{lm}$ for a {\it
non-uniform} rotation.  

There was a brief flurry in the 1990s of
searches for a rotation with a dipole dependence on position
\cite{Carroll:1997tc,Leahy:1997wj,Wardle:1997gu,Loredo:1997gn},
following a claimed
detection \cite{Nodland:1997cc}.  Here I revisit and update such
measurements and generalize to higher-$l$ moments.  I search the
recent UV data \cite{Alighieri:2010eu}, combined with the radio
constraint from 3C 9 \cite{Leahy:1997wj}, and find no evidence for any
rotation with a dipole or quadrupole dependence on position.
I constrain the $\alpha_{lm}$ (for {\it any} $l$) to
$(4\pi)^{-1/2} \alpha_{lm} \lesssim 3.7^\circ$ ($1\sigma$), and I
place a constraint to the power spectrum
for $\alpha$ for theories that predict
that $\alpha(\theta,\phi)$ is a stochastic field.  To preface, I
discuss the derivation of the constraint, $\alpha =-0.7^\circ \pm
2.0^\circ$ (to redshifts $z\simeq2$), from the UV data, a result 
that is strengthened to $\alpha = -0.1^\circ\pm
1.7^\circ$ if the radio data on 3C 9 is included.
Finally, I discuss how the analysis of high-resolution
intensity-polarization maps of individual sources can be
optimized, using tensor-harmonic techniques similar to those for
CMB polarization and weak gravitational lensing.

{\it Prelude: A uniform rotation.} Table \ref{tab:data}
reproduces data on 8 UV sources from Ref.~\cite{Alighieri:2010eu}
as well as radio data on 3C 9 \cite{Leahy:1997wj,Kronberg}.  Listed
there are the positions $(\theta_i,\phi_i)$,
position-angle--polarization offsets $\alpha_i$, and measurement
errors $\sigma_i$ to these offsets.

Let us first test with this data whether there is a rotation, by
an angle $\alpha$, that is uniform across the sky.  We will also
determine the scatter $\sigma_p$ in the measurements of $\alpha$.
To estimate the mean offset from the data, we use the
minimum-variance estimator,
\begin{equation}
     \widehat\alpha = \left.  \left[ \sum_i
     \alpha_i/\sigma_i^2 \right] \middle/ \left[ \sum_i
     1/\sigma_i^2 \right] \right. .
\label{eqn:uniformmean}
\end{equation}
The error to our measurement of $\alpha$ is then inverse root of
the denominator.  We find for the UV data in
Table~\ref{tab:data} $\alpha=-0.7\pm2.0$.  Including
the radio source 3C~9 improves the minimum-variance constraint
to $\alpha=-0.1^\circ\pm1.7^\circ$. In general, it could be that
the sample contains a significant intrinsic scatter in
the offsets, in which case the minimum-variance error would
underestimate the true
error in $\alpha$.  For example, one extremely
well-measured and statistically-significant nonzero offset might
suggest nonzero CB, but could alternatively be due to
an intrinsic offset in the source.  Still, the
measured offsets for the current sample are well within their
measurement errors, and so the error obtained here is
probably sound.  The error $\sigma_\alpha^2 = \sum_i
(\alpha_i-\widehat\alpha)^2/[N(N-1)]$ obtained from the measured
dispersion is in fact a bit smaller, $\sigma_\alpha=1.4^\circ$,
suggesting that the reported measurement errors in this sample
may be a bit high and the true constraint a bit stronger.

\begin{table}[t]
\begin{center}
\begin{tabular}{c|c|c|c|c}
i & $\theta_i$ (deg) & $\phi_i$ (deg) & $\alpha_i$
(deg) & $\sigma_i$ (deg) \\
\hline
1 & 78 & 34 & $-1.0$ & 3.5 \\
2 & 66 & 146 & $-0.3$ & 4.4 \\
3 & 109 & 128 & 1.6 & 4.5 \\
4 & 90 & 213 & $-8.0$ & 8.0 \\
5 & 93 & 191 & $-4.0$ & 8.8 \\
6 & 68 & 307 & $-4.0$ & 9.0 \\
7 & 114 & 317 & 4.6 & 9.7 \\
8 & 103 & 20 & 5.0 & 16 \\
{\it 9} & {\it 5} & {\it 105} & {\it 2} & {\it 3}\\
\end{tabular}
\end{center}
\caption{The $\theta_i$-$\phi_i$ coordinates, offsets $\alpha_i$
     and measurement errors $\sigma_i$ for the eight sources
     listed in Ref.~\protect\cite{Alighieri:2010eu} plus (the
     last) the radio source 3C 9 (from
     Ref.~\protect\cite{Leahy:1997wj}).}
\label{tab:data}
\end{table}

The scatter $\sigma_p$ in the offsets is determined from the data via,
\begin{equation}
     \sigma_p^2 = \left[\sum_i (\alpha_i -
     \widehat\alpha)^2\sigma_i^{-2} \right] \left[\sum_i
     \sigma_i^{-2} - \sum_i \sigma_i^{-4}/\sum_i\sigma_i^{-2}
     \right]^{-1} .
\label{eqn:scatter}
\end{equation}
The 9 sources in Table \ref{tab:data} result in
$\sigma_p=2.9^\circ$, a result that will be used below.
Note that the weighted estimate of the scatter in
Eq.~(\ref{eqn:scatter}) is a bit smaller than the value
$4.4^\circ$ obtained if an unweighted estimator, $\sigma_p^2 =
(N-1)^{-1} \sum_i (\alpha_i-\widehat\alpha)^2$, for the variance
is used, an indication that the unweighted variance in this data
is due primarily to measurement error, not intrinsic scatter.

{\it Non-uniform rotation.}  Now suppose we wish to test if
there is a single $lm$ spherical-harmonic variation in
$\alpha(\theta,\phi)$: i.e., that
$\alpha(\theta,\phi)=\alpha_{lm} Y_{lm}(\theta,\phi)$, for some
given $l$ and $m$.  Then each data point would provide an
estimator, $\widehat\alpha_{lm}^i =(\alpha_i
-\widehat\alpha)/Y_{lm}(\theta_i,\phi_i)$,
with variance $\VEV{ |\widehat\alpha_{lm}^i|^2} =
\sigma_i^2/ | Y_{lm}(\theta_i,\phi_i)|^2$.
I include the $\widehat\alpha$ term in the estimator to avoid confusing a higher
moment (e.g., a dipole) with uniform rotation in case of limited
or irregular sky coverage.  It should become irrelevant in an
ideal experiment, with $N\to\infty$ and a population of sources
spread uniformly throughout the sky.  Note that
$\widehat\alpha_{lm}^i$ is complex, and the variances to the real
and imaginary parts are each $\VEV{ |\widehat\alpha_{lm}^i|^2}/2$.

The minimum-variance estimator $\widehat\alpha_{lm}$ obtained
from all $N$ data points is obtained by adding all the $N$ individual
$\widehat\alpha_{lm}^i$ estimators with inverse-variance
weighting; i.e.,
\begin{equation}
     \widehat\alpha_{lm} = \left. \left[ \sum_i
     \frac{\alpha_i-\widehat\alpha}{\sigma_i^2}
     Y_{lm}^*(\theta_i,\phi_i)\right] \middle/ \left[ \sum_i
     \frac{ |Y_{lm}(\theta_i,\phi_i)|^2}{\sigma_i^2}
     \right] \right. ,
\label{eqn:alphalmestimator}
\end{equation}
with variance given by the inverse of the denominator in this
expression.  The results of such an
analysis of the 9 sources in Table \ref{tab:data} are presented
in Table~\ref{tab:alms}.  There is no evidence for any nonzero
$\alpha_{lm}$ for $l\leq2$.

\begin{table}[htbp]
\begin{center}
\begin{tabular}{c|c|c|c}
$l$ & $m$ & $(4\pi)^{-1/2}\widehat\alpha_{lm}$ (deg) &
$(4\pi)^{-1/2}\VEV{|\alpha_{lm}|^2}^{1/2}$ (deg) \\
\hline
1 & 0 & $-2.9$ & 3.4 \\
1 & 1 & $-0.7-0.3i$ & 1.4 \\
2 & 0 & 0.2 & 2.0\\
2 & 1 & $1.1+0.2i$ & 2.3\\
2 & 2 & $0.2-0.5i$ & 1.3\\
\end{tabular}
\end{center}
\caption{The measured $\alpha_{lm}$ obtained from the data in
     Table~\ref{tab:data}.}
\label{tab:alms}
\end{table}

The values of the individual $\widehat\alpha_{lm}$s, for a
given $l$, depend on the choice of coordinate system.  
To test for a non-uniform CB in a rotationally-invariant
way, one must evaluate the rotational invariants $\widehat
C_l= \sum_{m=-l}^l  |\widehat\alpha_{lm}|^2/ (2l+1)$.
Doing so, no evidence of a non-uniform CB is found for the
dipole ($l=1$) and quadrupole ($l=2$).  Rough
upper limits to the dipole and quadrupole amplitudes can be
obtained from the noise: $\sqrt{C_1/(4\pi)} \lesssim 2.3^\circ$
and $\sqrt{C_2/(4\pi)} \lesssim 1.9^\circ$.


{\it Higher-$l$ moments.} Since we have in the current analysis
only 8 data points, it is
not really possible to measure any $\alpha_{lm}$s with $l\gtrsim
2$.  However, if there were a
nonzero $\alpha_{lm}$ for some high $l$, it would give rise
to a scatter in the measured $\alpha_i$s with variance,
$\VEV{\alpha^2} = (4\pi)^{-1} \int \, d\hatn\, \left
     [\alpha(\theta,\phi) \right]^2 =
     |\alpha_{lm}^2|/(4\pi)$.
If the sources are randomly distributed on the sky at points
with angular separations $\Delta\theta \gg \pi/l$, then this
variance $\VEV{\alpha^2}$ cannot be larger than the variance in
the data.  The variance measured from the data in
Table~\ref{tab:data} is roughly $(2.9^\circ)^2$, but there is a
sample error to this variance, of roughly $\sqrt{2/N}$, where
$N$ is the number of (statistically significant) data points,
which I estimate to be 5.  I therefore take, as a rough upper
limit $\VEV{\alpha^2}^{1/2} \lesssim 3.7^\circ$, implying
$|\alpha_{lm}|^2/4\pi \lesssim (3.7^\circ)^2$.  The upper limit to any
$C_l$ are similar: $ C_l / 4\pi \lesssim (3.7^\circ)^2$ for any
individual higher $l$.

{\it Stochastic values of $\alpha$.}
Theories with a spatially-varying $\alpha(\theta,\phi)$ 
generally predict that $\alpha(\theta,\phi)$ is a realization of a
random field with some given power spectrum $C_l$.  Such a theory
results in a variance in the measured offsets of $\VEV{\alpha^2} =
\sum_l (2l+1)C_l/(4\pi)$
which, again, must be $\lesssim (3.7^\circ)^2$.  For example,
suppose some theory predicts a scale-invariant spectrum, $l(l+1)
C_l = 2 C_1 \exp(-l^2/l_c^2)$ from $l=1$ out to some cutoff
moment $l_c$ with an amplitude parametrized by the dipole moment
$C_1$.  Then approximating for this model
$\VEV{\alpha^2} \simeq C_1\ln(7 l_c) / \pi$, 
we find the dipole to be constrained to $C_1 /4\pi \lesssim
(1/4) \VEV{\alpha^2} /\ln(7l_c) \simeq (0.7^\circ)^2/\ln(l_c/200)$.
Note that the finite angular size $\delta$ of the images limits
the effective $l_c \lesssim200\, (\delta/1^\circ)^{-1}$, even if
the theory allows it to be much larger.  

{\it Analysis of individual objects.}
Early measurements of CB
\cite{Carroll:1989vb,Nodland:1997cc,Loredo:1997gn}
considered simply the offset between the image position angle
and the mean polarization averaged over the entire image.
But this averaging erases much of the
information available in the
source~\cite{Cimatti:1993yc,Wardle:1997gu,Leahy:1997wj}.  If a
high-resolution map
of the intensity and polarization of a given source is
available, then the offset between the intensity gradient and
the polarization {\it throughout a given source} can provide a
far more precise measurement of the mean offset for that
particular source.  The sensitivity to CB from a handful of
well-resolved sources can thus compete with that of hundreds of
unresolved sources.

Still, one can do better in terms of measuring the offset
$\alpha$ from a given well-resolved source than prior analyses by
using techniques developed to quantify temperature-polarization
correlation functions in the CMB
\cite{Kamionkowski:1996ks,Zaldarriaga:1996}, and also
galaxy-shape correlations induced by weak gravitational lensing
\cite{Stebbins:1996wx}.  These techniques deal, for example,
with the ambiguity in the direction of the linear polarization
and also with optimizing low signal-to-noise measurements.  They
allow the full two-point intensity-polarization correlations to
be used, rather than simply the correlations at zero lag (as prior
analyses have used).  They provide additional information on
$\alpha$ from the polarization alone, even without
cross-correlation with the intensity, something that could not
be done with the more heuristic intensity-gradient--polarization
correlation.  These techniques are analogous to those for
measuring a uniform rotation angle with the CMB \cite{Lue:1998mq}.

Suppose that we have a resolved map of the intensity
$I(\theta_x,\theta_y)$ and Stokes parameters
$Q(\theta_x,\theta_y)$ and $U(\theta_x,\theta_y)$ of a given
radio source, where $\theta_x$ and $\theta_y$ are coordinates in
the image plane.  One first Fourier transforms, $\tilde
I(l_x,l_y) = \int d^2\theta\, e^{i\vec l\cdot \vec \theta}
I(\vec\theta)$, and similarly for $\tilde Q(\vec l)$ and $\tilde
U(\vec l)$, from which are obtained the rotational invariants,
\begin{eqnarray}
     \tilde E(\vec l) &=& \frac{1}{2} \frac{ (l_x^2-l_y^2) \tilde
     Q(\vec l) +  2 l_x l_y \tilde U(\vec l)}{l_x^2+l_y^2},
     \nonumber \\
     \tilde B(\vec l) &=& \frac{1}{2} \frac{ 2l_x l_y \tilde
     Q(\vec l) - (l_x^2-l_y^2) \tilde U(\vec l)}{l_x^2+l_y^2}.
\label{eqn:EandB}
\end{eqnarray}
From these, the six power spectra $P_l^{XX'} = \langle \tilde
X(\vec l) \tilde X'(\vec l)^* \rangle$ (where the angle brackets
denote an ensemble average) can be obtained, where
$\{X,X'\} = \{I, E,B\}$.  The B modes have opposite parity from
the I and E modes, and so we should have $P_l^{IB}=P_l^{EB}=0$,
if parity is preserved.  While any given source may in principle
have some handedness, and thus possibly nonzero IB or EB
correlations, there should be no preference for a given
handedness when averaging over many sources.  More importantly,
the existing measurements, which show that the offset $\alpha$
is small in the sources where it is measured, suggest that IB
and EB correlations will be small.

If CB rotates the polarization by an
angle $\alpha$, then part of the E mode is rotated into a B
mode, $\delta \tilde B(\vec l) =  \tilde E(\vec l) \sin2\alpha
\simeq 2 \alpha \tilde E(\vec l)$, thus inducing nonzero
$P_l^{IB}  = 2\alpha P_l^{IE}$ and $P_l^{EB} = 2 \alpha P_l^{EE}$.
A rotation-angle estimate is then obtained by comparing the
measured IB and EB correlations with the measured IE and EE
correlations, respectively.  

A similar analysis can be done, alternatively and equivalently,
using temperature-polarization two-point correlation functions
\cite{Kamionkowski:1996ks,Stebbins:1996wx}.  This involves
taking all pairs $(\vec\theta_1,\vec\theta_2)$ of points in the
map, and then measuring correlations
between the intensity $I$ and Stokes parameters $Q_r$ and $U_r$
measured in a coordinate system that is aligned with the line
connecting the two points.  Again, symmetry considerations
suggest that, in the absence of CB,
$\langle{I(\vec\theta_1) U_r(\vec\theta_2)}\rangle =
\langle{Q_r(\vec\theta_1) U_r(\vec\theta_2)}\rangle=0$.  If the
polarizations are rotated by an angle $\alpha$, then these
parity-odd correlations are induced, with magnitudes
$\langle{I(\vec\theta_1) U_r(\vec\theta_2)}\rangle = 2\alpha \langle{
I(\vec\theta_1) Q_r(\vec\theta_2)}\rangle$ and 
$\langle{Q_r(\vec\theta_1) U_r(\vec\theta_2)}\rangle = 2\alpha \langle{
Q_r(\vec\theta_1) Q_r(\vec\theta_2)}\rangle$.  The decision as to
whether to use power spectra or correlation functions will
depend on the noise properties of the map.

{\it Discussion.}
Here I have discussed measurements of a
CB rotation of the linear polarization
that varies as a function of position on the sky and derived
rough upper limits to rotation-angle multipole moments and power
spectra. I discussed how the analysis of future high-resolution
intensity-polarization images of high-redshift sources can be
optimized with techniques analogous to those in CMB-polarization
studies.

The analysis presented here is meant primarily to be
illustrative.  The existing data are far from optimized for this
particular measurement.  First of all, I used only 9 sources at
redshifts $z\gtrsim 2$, and the statistical weight is dominated
by only half of those.  Moreover, they are not uniformly spread
on the sky (which is why the errors to the different $m$ moments
for a given $l$ vary so widely), and this could give rise to
pitfalls.  Had my analysis found evidence for a signal, this may
have been cause for concern. But given that the results are
null, the derived upper limits are probably sound.

Although a comparable sensitivity to a position-dependent
rotation can in principle be obtained from existing CMB data,
the analysis is difficult and has not yet been done.  The
simple exercise I have performed here is thus the strongest
existing constraint to a position-dependent rotation, at least
for a rotation that occurs at redshifts $0<z\lesssim 2$.  A
model that predicts rotation at $3<z\lesssim1100$ could still
produce a signal in the CMB without violating the constraint I
have derived.  Likewise, a slightly stronger constraint can
probably be obtained from the radio-galaxy data in
Ref.~\cite{Leahy:1997wj}, although for lower-redshift sources,
and thus over a smaller baseline.  Whether that constraint would
be competitive with the one I have derived would, again, depend
on the redshift dependence of $\alpha$ in any given model.  Of
course, if one has a particular model that makes a specific
prediction for the redshift dependence of $\alpha(\hatn;z)$,
then the data from sources at all redshifts can be combined to
provide optimal estimators for the parameters of the model.

Improvements to the illustrative analysis I have done here
should be straightforward.  The error to the $\alpha_{lm}$
should scale simply as $N^{-1/2}$ with the number $N$ of
sources, assuming similar image qualities to those obtained so
far.  Thus, for example, if the sample I used of $N\sim4$
well-measured offsets can be improved to $N\sim400$, the
sensitivity will be competitive with that expected from Planck.
However, progress can be accelerated, beyond $N^{-1/2}$, if
more precise offset measurements can be obtained for at least
some of these individual sources, either from better images, an
improved analysis, or both.  The goal of identifying the new
physics responsible for cosmic acceleration will hopefully
motivate such empirical investigations.

\begin{acknowledgments}
I thank S.\ di Serego Alighieri, R.\ Caldwell, V.\ Gluscevic,
and A.\ Readhead for useful discussions.  This work was supported
by DoE DE-FG03-92-ER40701, NASA NNX10AD04G, and the Gordon and
Betty Moore Foundation.
\end{acknowledgments}


\begin{thebibliography}{}


\bibitem{darkenergy}
  E.~J.~Copeland, M.~Sami and S.~Tsujikawa,
  Int.\ J.\ Mod.\ Phys.\  D {\bf 15}, 1753 (2006)
  [arXiv:hep-th/0603057];
  R.~R.~Caldwell and M.~Kamionkowski,
  Ann.\ Rev.\ Nucl.\ Part.\ Sci.\  {\bf 59}, 397 (2009)
  [arXiv:0903.0866 [astro-ph.CO]];
  A.~Silvestri and M.~Trodden,
  Rept.\ Prog.\ Phys.\  {\bf 72}, 096901 (2009)
  [arXiv:0904.0024 [astro-ph.CO]];
  A.~J.~Albrecht {\it et al.},
  arXiv:astro-ph/0609591;
  E.~V.~Linder,
  Rept.\ Prog.\ Phys.\  {\bf 71}, 056901 (2008)
  [arXiv:0801.2968 [astro-ph]];
  J.~Frieman, M.~Turner and D.~Huterer,
  Ann.\ Rev.\ Astron.\ Astrophys.\  {\bf 46}, 385 (2008)
  [arXiv:0803.0982 [astro-ph]].

\bibitem{quintessence}
  B.~Ratra and P.~J.~E.~Peebles,
  Phys.\ Rev.\  D {\bf 37}, 3406 (1988);
  C.~Wetterich,
  Astron.\ Astrophys.\  {\bf 301}, 321 (1995)
  [arXiv:hep-th/9408025];
  K.~Coble, S.~Dodelson and J.~A.~Frieman,
  Phys.\ Rev.\  D {\bf 55}, 1851 (1997)
  [arXiv:astro-ph/9608122];
  M.~S.~Turner and M.~J.~White,
  Phys.\ Rev.\  D {\bf 56}, 4439 (1997)
  [arXiv:astro-ph/9701138];
  R.~R.~Caldwell, R.~Dave and P.~J.~Steinhardt,
  Phys.\ Rev.\ Lett.\  {\bf 80}, 1582 (1998)
  [arXiv:astro-ph/9708069].

\bibitem{Carroll:1998zi}
  S.~M.~Carroll,
  Phys.\ Rev.\ Lett.\  {\bf 81}, 3067 (1998)
  [arXiv:astro-ph/9806099].

\bibitem{Carroll:1989vb}
  S.~M.~Carroll, G.~B.~Field and R.~Jackiw,
  Phys.\ Rev.\  D {\bf 41}, 1231 (1990).

\bibitem{Lue:1998mq}
  A.~Lue, L.~M.~Wang and M.~Kamionkowski,
  Phys.\ Rev.\ Lett.\  {\bf 83}, 1506 (1999)
  [arXiv:astro-ph/9812088];
  N.~F.~Lepora,
  arXiv:gr-qc/9812077.

\bibitem{Feng:2006dp}
  B.~Feng {\it et al.}, 
  Phys.\ Rev.\ Lett.\  {\bf 96}, 221302 (2006)
  [arXiv:astro-ph/0601095];
 P.~Cabella, P.~Natoli and J.~Silk,
  Phys.\ Rev.\  D {\bf 76}, 123014 (2007)
  [arXiv:0705.0810 [astro-ph]];
  T.~Kahniashvili, R.~Durrer and Y.~Maravin,
  arXiv:0807.2593 [astro-ph];
  J.~Q.~Xia {\it et al.}, 
  arXiv:0803.2350 [astro-ph];
  J.~Q.~Xia {\it et al.}, 
  arXiv:0710.3325 [hep-ph];
 :.~E.~Y.~Wu {\it et al.}  [QUaD Collaboration],
 arXiv:0811.0618 [astro-ph];
  L.~Pagano {\it et al.},
  arXiv:0905.1651 [astro-ph.CO];
  E.~Komatsu {\it et al.},
  arXiv:1001.4538 [astro-ph.CO].

\bibitem{Pospelov:2008gg}
  M.~Pospelov, A.~Ritz and C.~Skordis,
  arXiv:0808.0673 [astro-ph];
  M.~Li and X.~Zhang,
  arXiv:0810.0403 [astro-ph];
  R.~R.~Caldwell, in preparation.

\bibitem{Gardner:2006za}
  S.~Gardner,
  Phys.\ Rev.\ Lett.\  {\bf 100}, 041303 (2008)
  [arXiv:astro-ph/0611684].

\bibitem{Kamionkowski:2008fp}
  M.~Kamionkowski,
  Phys.\ Rev.\ Lett.\  {\bf 102}, 111302 (2009)
  [arXiv:0810.1286 [astro-ph]];
  V.~Gluscevic, M.~Kamionkowski and A.~Cooray,
  Phys.\ Rev.\  D {\bf 80}, 023510 (2009)
  [arXiv:0905.1687 [astro-ph.CO]];
  A.~P.~S.~Yadav {\it et al.},
  Phys.\ Rev.\  D {\bf 79}, 123009 (2009)
  [arXiv:0902.4466 [astro-ph.CO]].

\bibitem{Nodland:1997cc}
  B.~Nodland and J.~P.~Ralston,
  Phys.\ Rev.\ Lett.\  {\bf 78}, 3043 (1997)
  [arXiv:astro-ph/9704196].

\bibitem{Loredo:1997gn}
  T.~J.~Loredo, E.~E.~Flanagan and I.~M.~Wasserman,
  Phys.\ Rev.\  D {\bf 56}, 7507 (1997)
  [arXiv:astro-ph/9706258];
  D.~J.~Eisenstein and E.~F.~Bunn,
  Phys.\ Rev.\ Lett.\  {\bf 79}, 1957 (1997)
  [arXiv:astro-ph/9704247].

\bibitem{Wardle:1997gu}
  J.~F.~C.~Wardle, R.~A.~Perley and M.~H.~Cohen,
  Phys.\ Rev.\ Lett.\  {\bf 79}, 1801 (1997)
  [arXiv:astro-ph/9705142].

\bibitem{Leahy:1997wj}
  J.~P.~Leahy,
  arXiv:astro-ph/9704285.

\bibitem{Carroll:1997tc}
  S.~M.~Carroll and G.~B.~Field,
  Phys.\ Rev.\ Lett.\  {\bf 79}, 2394 (1997)
  [arXiv:astro-ph/9704263].

\bibitem{Cimatti:1993yc}
  A.~Cimatti, S.~di Serego Alighieri, G.~B.~Field and R.~A.~E.~Fosbury,
  Astrophys.\ J.\  {\bf 422}, 562 (1994).

\bibitem{Alighieri:2010eu}
  S.~d.~S.~Alighieri, F.~Finelli and M.~Galaverni,
  arXiv:1003.4823 [astro-ph.CO].

\bibitem{Kronberg}
  P.~P.~Kronberg, C.~C.~Dyer, and H.-J.~R\"oser,
  Astrophys.\ J.\ {\bf 472}, 115 (1996).

\bibitem{Kamionkowski:1996ks}
  M.~Kamionkowski, A.~Kosowsky and A.~Stebbins,
  Phys.\ Rev.\  D {\bf 55}, 7368 (1997)
  [arXiv:astro-ph/9611125];
  Phys.\ Rev.\ Lett.\  {\bf 78}, 2058 (1997)
  [arXiv:astro-ph/9609132];
  P.~Cabella and M.~Kamionkowski,
  arXiv:astro-ph/0403392.


\bibitem{Zaldarriaga:1996}
  M.~Zaldarriaga and U.~Seljak,
  Phys.\ Rev.\  D {\bf 55}, 1830 (1997)
  [arXiv:astro-ph/9609170].
  U.~Seljak and M.~Zaldarriaga,
  Phys.\ Rev.\ Lett.\  {\bf 78}, 2054 (1997)
  [arXiv:astro-ph/9609169].

\bibitem{Stebbins:1996wx}
  A.~Stebbins,
  arXiv:astro-ph/9609149;
  M.~Kamionkowski {\it et al.},
  Mon.\ Not.\ Roy.\ Astron.\ Soc.\  {\bf 301}, 1064 (1998)
  [arXiv:astro-ph/9712030].

\end{thebibliography}
\end{document}